\documentclass[runningheads]{llncs}
\usepackage{amssymb}
\usepackage{amsmath}
\usepackage{color}
\begin{document}
	\title{Zero-Knowledge Proof in NuLink}
	\author{ Pawn\inst{1}  \and Rookie\inst{1} \and
		Zhuan Cheng\inst{1}$^{*}$ 
	}
	\authorrunning{Pawn et al.}
	
	\institute{$^{*}$NuLink Network Foundation
	}
	
\maketitle

 \begin{abstract}
     NuLink provides privacy-preserving technology for decentralized applications via APIs. Users can securely store its valuable data, trade with others and so on. To ensure the privacy and security of service provided by NuLink, (zero-knowledge) proof systems are necessary. 

     Zero-knowledge proof systems allow the prover to make the verifier believe that a certain conclusion is correct without providing any useful information to the verifier. In NuLink, we are going to use (zero-knowledge) proof system in the following three methods:

     \begin{enumerate}
         \item Users store their data through NuLink in a decentralized manner. To ensure that the storage clients are indeed storing the data, we employ proof of storage systems. In this system, users prepare certain challenges that can only be correctly answered by those who are actually storing the data.
         \item Users have the option to outsource computations to NuLink. To verify the correctness of the computation results provided by the compute node, we require the node to provide a proof of correctness via SNARK systems. When sensitive parameters are used as inputs for computation, we utilize zk-SNARKs to prevent any potential leakage of these parameters.
         \item Users may choose to trade their data through NuLink. To confirm that the buyer has sufficient digital funds and the seller possesses the desired data, both parties can provide a proof via zk-SNARKs. This builds confidence and prevents cheating during transactions.
     \end{enumerate}

     Using zero-knowledge proof systems, we can ensure that all nodes in NuLink behaves honestly and avoid cheating in the whole system.

 \end{abstract}
	\section{Introduction}
With the advancement of information technology, a plethora of online services have emerged, including e-commerce, digital currency, and outsourced computation. These services offer immense convenience to users. For instance, a user can outsource their data (e.g., a song they created) to online nodes, outsource the computation (e.g., the creator might commission someone to modify their song) of their data online, and even conduct online trades (e.g., the creator might sell their song to others).

 However, the potential for malicious behavior by service providers or users exists. For instance, a storage or computation node might attempt to access the sensitive data of an honest user. On the other hand, a storage or computation node might neglect to store the data or carry out the computation to save resources, or worse, they might tamper with the stored or computed data for their own benefit. These malicious actions can result in losses for honest users.

 Addressing these privacy and security issues is the driving force behind NuLink\cite{NuLink}. NuLink possesses several core characteristics: it integrates a variety of cryptographic technologies, operates in a decentralized manner, is easy to implement, and is open source. Our goal is to provide a ready-to-use solution that reduces the barriers to implementing a privacy protection scheme in applications for all types of businesses. NuLink will provide everything needed, including data encryption, key and storage management, inter-blockchain deployment, and privacy computing.

 In this paper, we will demonstrate how to use zero-knowledge proof to foster trust between clients and nodes without exposing sensitive information.

	\section{What is Zero-Knowledge Proof}
 Zero-Knowledge Proof (ZKP) \cite{GMR89} allows the prover makes the verifier believe that a certain conclusion is correct without providing any useful information to the verifier. For more details, Zero-Knowledge Proof for an NP language $L$ is a protocol between a prover $P$ and a verifier $V$. At the beginning of protocol, $P$ takes the statement $x\in L$ and corresponding witness $w\in R_L(x)$ as input and $V$ only takes $x$ as input; after the execution of protocol, $V$ should only learn the truth that ``$x\in L$'' but nothing else.  To present a formal definition of zero-knowledge proofs, we first introduce proof protocol:
\begin{definition}[Proof Protocol] A protocol $\langle P,V \rangle$ for NP relation $R_L(x,w)$ is a proof protocol if it satisfies:
\begin{enumerate}
\item \textbf{Completeness}: For any $n \in \mathbb{N}, x\in L\cap \{0,1\}^n,w\in R_L(x)$:
$$\emph{\textsf{Out}}_V\langle P(w),V\rangle(x)=1$$
\item \textbf{Soundness}: For any malicious prover $P^*=\{P^*_n\}_{n\in \mathbb{N}}$, there exists a negligible function $negl$ such that for any security parameter $n\in \mathbb{N}$ and any $x\in \{0,1\}^n\backslash L$,
$$\Pr[\emph{\textsf{Out}}_V\langle P_n^*,V\rangle (x)=1]<negl(n)$$
\end{enumerate}
\end{definition}
An argument protocol is a weak version of proof protocol where the soundness security only holds for polynomial-size malicious prover.

Then, we give the definition of zero-knowledge:
\begin{definition}[Zero-Knowledge] We call a proof/argument protocol $\langle P,V \rangle $ is zero-knowledge if for any polynomial-sized malicious verifier $V^*=\{V^*_n\}_{n\in \mathbb{N}}$, there exists a polynomial-sized simulator ${{S}}=\{{{S}}_n\}_{n\in\mathbb{N}}$ such that for any polynomial-sized distinguisher ${{D}}=\{{{D}}_n\}_{n\in \mathbb{N}}$ there exists a negligible function $negl$ such that for any security parameter $n\in\mathbb{N}$, and any statement $x\in L\cap\{0,1\}^n,\,w\in R_L(x)$:
$$\left|\Pr\left[{{D}}_n(\emph{\textsf{View}}_{V_n^*}^{P(w)}(x))=1\right]-\Pr\left[{{D}}_n({{S}}_n(x))=1\right]\right|<negl(n)$$
The probability is over the coins of ${{D}}_n,V_n^*,P,{{S}}_n$.
\end{definition}

 Zero-Knowledge Proof was first proposed by S Goldwasser et al in 1989\cite{GMR89}. Since the the first construction of zero-knowledge proof for every NP language put forward by Goldreich in 1991 \cite{GMW91}, zero-knowledge proof have become a powerful tool in cryptography: it provides a key tool for the construction of multiparty secure computation -- the universal solution for almost all cryptography task. Furthermore, zero-knowledge protocol are widely used in lots of special-purpose cryptography protocols, e.g., identification, digital money, digital vote, group signature. 

 Due to the wide usage of zero-knowledge protocol, various constructions are put forward. In the plain model, we knows how to construct zero-knowledge argument in four-round\cite{GK96}, and recently, several works shows how to construct three-round distinguisher-dependent simulatable zero-knowledge protocols from standard assumptions\cite{JKKR17,BKP19}. 
 
 However, several negative results shows that, in the plain model, it is impossible to construct zero-knowledge argument in two-round \cite{GO94}, and impossible to construct zero-knowledge argument with black-box simulator in three-round\cite{GK96}. Even though there are lots of studies of non-black-box simulation technique \cite{DGS09,CLP13,Goyal13,BP15,CPS13}, none of them achieves standard zero-knowledge argument in three rounds from standard assumptions. High round-complexity means that using these zero-knowledge argument would have a huge communication latency. Further, most of zero-knowledge protocol for every NP language in the plain model have a huge verification cost, which limits its usage.

 In practical, we usually use non-interactive arguments in the CRS (common reference string) model\cite{BFM88}. 
Compared with the plain model, the CRS model allows both parties have a common reference string as additional input. For more details, a non-interactive zero-knowledge argument in the CRS model consists of four algorithms $(Setup, Prove, Verify, Sim)$ such that:
\medskip

\noindent$(crs,td)\gets Setup(R_L)$: The setup produces a common reference string $crs$ and a simulation trapdoor $td$ for the relation $R_L$.

\noindent$\pi \gets Prove(R, crs, x, w)$: The prover algorithm takes  a common reference string $crs$ and $(x,w)\in R_L$ as input, and returns an argument $\pi$.

\noindent $0/1 \gets Verify(R, crs, x, \pi)$: The verification algorithm takes a common reference string $crs$, a statement $x$ and an argument $pi$ as input, and returns 0 (reject) or 1 (accept).

\noindent $\pi \gets Sim(R, td, x)$: The simulator takes as input a simulation trapdoor and statement $x$ and returns a simulation argument $\pi$.
\medskip

Similar with zero-knowledge argument in the plain model, non-interactive zero-knowledge argument also should satisfy the completeness, soundness and zero-knowledge properties. A formal definition is shown as follows:

\begin{definition}[Non-Interactive Zero-Knowledge Argument Scheme] A non-interactive zero-knowledge argument scheme consists of following four algorithms $(Setup, Prove, Verify, Sim)$ described above and satisfies that :
\begin{enumerate}
    \item \textbf{Completeness}: For any $n \in \mathbb{N}, x\in L\cap \{0,1\}^n,w\in R_L(x)$:
$$\Pr\left[\begin{gathered}(crs,td)\gets Setup(R_L),\\
\pi \gets Prove(R, crs, x, w)\end{gathered}: Verify(R, crs, x, \pi)=1\right]=1$$
\item \textbf{Soundness}: For any malicious prover $\mathcal{A}$, there exists a negligible function $negl$ such that for any security parameter $n\in \mathbb{N}$ and any $x\in \{0,1\}^n\backslash L$,
$$\Pr\left[\begin{gathered}(crs,td)\gets Setup(R_L),\\
\pi \gets \mathcal{A}(R, crs, x)\end{gathered}: Verify(R, crs, x, \pi)=1\right]<negl$$
\item  \textbf{Zero-Knowledge}: for any security parameter $n\in\mathbb{N}$, any statement $x\in L\cap\{0,1\}^n,\,w\in R_L(x)$ and any (polynomial-sized) adversary $\mathcal{A}$, the difference of the following two probability is negligible:
$$\begin{gathered}\Pr\left[\begin{gathered}(crs,td)\gets Setup(R_L),\\
\pi \gets Prove(R, crs, x, w)\end{gathered}: \mathcal{A}(R, crs, x, \pi)=1\right]\\
\Pr\left[\begin{gathered}(crs,td)\gets Setup(R_L),\\
\pi \gets Sim(R, td, x)\end{gathered}: \mathcal{A}(R, crs, x, \pi)=1\right]
\end{gathered}$$
\end{enumerate}
\end{definition}

Sometimes, we might consider a stronger version of soundness. That's knowledge soundness, which requires that: There exists an extractor such that for any PPT adversary generating an acceptable proof for statement $x$ with noticeable probability, the extractor can extract a valid witness $w\in R_L(x)$ from adversary with a probability negligibly close to 1. An argument satisfying knowledge soundness is called an argument of knowledge.

There has been a lot of progress both in theory and practice on constructing highly efficient non-interactive arguments with small size and low verification complexity as well as its zero-knowledge version, so-called (zero-knowledge) succinct non-interactive arguments, (zk-)SNARGs and (zero-knowledge) succinct non-interactive arguments of knowledge, (zk-)SNARKs.

Usually, SNARGs or SNARKs are constructed for proving circuit satisfiability problem or Rank-1 constrain satisfiability problem, and ``succinct'' requires that: 1) The prover run time is quasilinear in circuit size; 2) The proof length is logarithmic in circuit size; 3) The verifier run time is polylogarithmic in circuit size.

There are two main paths constructing practical zk-SNARKs, LPCP-based zk-SNARKs and PIOP-based zk-SNARKs. 

LPCP-based zk-SNARKs (e.g., \cite{GGPR13,Groth16}) follows the paradigm of ``Linear PCP system + Linear-only encryption''. The prover needs to transform the statement and witness into a proof of the Linear PCP system and the Setup algorithm encrypts random challenge matrix using the linear-only encryption. Finally, the proof generates the proof of zk-SNARKs scheme via doing linear operation over the ciphertexts according the LPCP proof. The most famous LPCP-based zk-SNARKs is Groth16 \cite{Groth16}, put forward by Jens Groth in 2016. Groth16 is designed for arithmetic circuit satisfiability and its proof consists of only 3 group elements. In addition to being small, the proof is also easy to verify. The verifier just needs to compute a number of exponentiations proportional to the statement size and check a single pairing product equation, which only has 3 pairings.  So far, Groth16 still has the one of the shortest proof size and fastest verification. We are going to use Groth16 scheme in NuLink in suitable situations.

However, Groth16 scheme as well as other LPCP-based zk-SNARKs schemes requires a trusted third party to generate the crs. The setup algorithm needs the description of language as input. This issue makes that Groth16 scheme is not friendly to be used in a decentralized situation.

PIOP-based zk-SNARKs (e.g., \cite{Setty20,GWC19,CHMMVW20}) follows the paradigm of ``polynomial commitment + polynomial interactive oracle proof (PIOP) system''. PIOP system is an interactive proof system except that in some step of prover, it will provide an oracle of some polynomial and the verifier can query the value of this polynomial on some points. To construct the PIOP-based zk-SNARKs, first require both parties to transform the description of language into a public-coin PIOP system, the prover of zk-SNARKs uses the witness to act as the prover of PIOP system and provide the polynomial oracle, and the verifier of zk-SNARKs acts as the verifier of PIOP system to check the proof. Secondly, replace the oracle of polynomial with a polynomial commitment. When the prover needs to provide the polynomial oracle, it generates a commitment of this polynomial; when the verifier wants to query the polynomial, it sends the query points to prover and prover open the polynomial commitment at these points. Finally, use the Fiat-Shamir heuristics to transform above interactive argument system into a non-interactive one. 

Most of PIOP-based zk-SNARKs don't need a trusted third party. Instead, PIOP-based zk-SNARKs work in the updatable and universal common reference strings model or common random string model. In the first model, a universal crs can be update many times and as long as there are at least one honest party who updates the crs or generate the crs, the crs is secure. And in the second model, one could use the hash of the statement or use a global random string as crs. 

As we have shown, the core of the construction of PIOP-based zk-SNARKs is of two folds: the choice of polynomial commitment and the approaches to generate the PIOP system. 

The are several mainstream polynomial commitment schemes. The most famous one is the KZG polynomial commitment\cite{KateZG10}, which is based on pairing-based groups. For the univariate version, the evaluation proof only consists constant group elements, the prover time is linear to the degree of polynomial and the verification only consists of constant pairing and exponentiation operations. The multivariate version\cite{ZGKPP17} also have great performance. One drawback of KZG-based scheme is that it requires a trusted setup and therefore most of zk-SNARKs scheme which uses KZG-based scheme is in the updatable and universal CRS model. Besides KZG-based polynomial commitment, there are lots of other schemes. We will show them as follows. 

 DL-based polynomial commitment: Bulletproof scheme is the most famous DL-based polynomial commitment\cite{BCCGP16,BBBPWM18}. This polynomial commitment has logarithmic evaluation proofs with great constants. Unfortunately, the verifier time is linear in the size of the polynomial. 

 FRI-based polynomial commitment. This series of polynomial schemes such as \cite{BBHR18,BGKS20,ZhangX0S20,KPV19} use
Reed-Solomon codeword to commit the polynomial and use the FRI protocol to generate evaluation proof. FRI-based polynomial commitment scheme have evaluation proofs of size and verifier time $O(\lambda \log_2(d))$ where $\lambda$ is the security parameter and $d = deg(f)$. 

 Tensor code-based polynomial commitment. This series of polynomial schemes \cite{XZS22,GLSTW23} use Tensor code to commit the polynomial. The prover time is strictly linear, that is, $O(2\mu)$ field operations and hashes where $\mu$ is the number of variables. The verifier time and proof size is $O_\lambda(\mu^2)$, which also improves the state-of-the-art. However, the concrete proof size is still unsatisfactory, e.g., for $\mu = 27$, the proof size is $6$ MBs. 

As for the approaches to PIOP system, there are also several famous approaches. We only introduce some of them as example.

GKR: GKR protocol \cite{GKR08} is presented by Goldwasser, Kalai, and Rothblum. It presents a approach to genrate the PIOP system. This protocol is best presented in terms of the (arithmetic) circuit evaluation problem. In this problem, Verifier and Prover first agree on a log-space uniform arithmetic circuit $C$ of fan-in 2 over a finite field $F$, and the goal is to compute the value of the output gate(s) of $C$. Letting $S$ denote the size (i.e., number of gates) of $C$ and $d$ denote the depth of $C$.  The costs to the verifier in the origin GKR protocol is $O(d \log S)$, which is sublinear to the size of $C$. GKR approach are suitable for low-depth circuit and not suitable for hight-depth one.

Spartan: Spartan \cite{Setty20} is presented by Srinath Setty. Spartan shows a method to transform the R1CS to the PIOP system. It
compactly encodes of an R1CS instance as a low-degree polynomial, and use sum-check protocol to prove the correctness of statement. With proper multilinear polynomial commitment, it could achieve linear prover time and logarithm verification time and communication. 

PLONK: PLONK scheme \cite{GWC19} is presented by Ariel Gabizon, Zachary J. Williamson, and Oana Ciobotaru. Among the IOP-based SNARKs that use a Polynomial-IOP, the PLONK system has emerged as one of the most widely adopted in industry. This is because PLONK proofs are very short (about 400 bytes in practice) and fast to verify. Moreover, Plonk supports
custom gates, could further improve the performance. The only issue is that the prover time of plonk is little high (or a circuit $C$ with $s$ arithmetic gates, the prover time is $O(s\log s)$). Luckily, there are various work \cite{CBBZ23} focus on this these question and achieves lots of positive results. 

By combine the polynomial commitments and PIOP systems, we can obtain (zk-)SNARKs systems. For example, ``Halo 2''\cite{halo,halo2}  combines the PlonK constant-round polynomial IOP with the Bulletproofs polynomial commitment scheme. ``PlonKy2''\cite{plonky2} uses a FRI-based polynomial commitment scheme rather than Bulletproofs.

Among these works, there are several schemes which perform well and have been used in lots of blockschain systems. For example, PLONK and its developing scheme (i.e., PlonKy2 developed by ploygon and halo2 developed by ZCash), has been used in ZKRollup, ZKEVM and other applications. We are going to use PLONK scheme and its developing scheme in NuLink in suitable situations due to its decentralized property and supporting custom gate.

	\section{What is NuLink}

 NuLink network\cite{NuLink} is a decentralized solution for web3 privacy-preserving applications developers to implement best practices and best of breed security and privacy. The NuLink platform provides endpoint encryption and cryptographic access control. Sensitive user data can be securely shared from any user platform to cloud or decentralized storage, and access to that data is granted automatically by policy in Proxy Re-Encryption or Attribute-Based Encryption. To verify the data source, data users can utilize Zero-Knowledge Proof. Additionally, for advanced privacy-preserving use cases, NuLink utilizes Fully Homomorphic Encryption (FHE) to provide customized enterprise-level data computation services. 

The NuLink network integrates the Application Layer, the Cryptograph Layer, the Storage Layer, the Blockchain Layer and the Watcher Network. 

\begin{enumerate}
    \item The Application Layer: The Application Layer acts as an interface between the system and the application, facilitating direct communication with the application while also liaising with the Cryptography Layer to validate access to the application's confidential information. 
    \item The Cryptograph Layer: The Cryptography Layer carries out cryptographic functions for the Application Layer, such as generating keys, encrypting, decrypting, and other related tasks. It also connects to the Storage Layer to facilitate the uploading and downloading of encrypted privacy data. 
    \item The Storage Layer: Our platform's Storage Layer is a secure network created for the purpose of storing confidential data in encrypted form. At present, we utilize IPFS (InterPlanetary File System) as the primary decentralized storage network. Nonetheless, we intend to incorporate additional storage networks like S3 in the coming times.
    \item The Blockchain Layer: The Blockchain Layer is responsible for managing staking node registration and service requests within the blockchain system. As of now, only Ethereum is supported for staking node registration. Nevertheless, users can still make service requests in other blockchain systems, such as Binance Smart Chain, Polygon, Polkadot, Arbitrum, Aptos or Sui.
    \item The Watcher Network: The Watcher Network is a relayer network that transfers staking node information from Ethereum to other blockchain systems. To ensure its decentralization and security, the Watcher Network is maintained under an on-chain governance mechanism (DAO). 
\end{enumerate}

Through a unified API integration, NuLink users can access a range of privacy-preserving services, storage, and blockchain solutions. Staking nodes have the opportunity to earn NuLink's token (NLK) by offering privacy-enhancing services in the Cryptography Layer, providing decentralized storage solutions in the Storage Layer, and relaying data from Ethereum in the Watcher Network

There are several example application scenarios, such as:
\begin{itemize}
\item Encrypted NFTs Trading Market
\item Privacy-Preserving Social Network
\item Decentralized Digital Rights Management
\item Electronic Health Records Sharing
\item Automotive Data Sharing
\end{itemize}

Within the NuLink network, Zero-Knowledge Proof is used to ensure that all functional nodes, including storage nodes, computing nodes, and proxy nodes, have publicly verifiable data processing and computing operations. At the same time, prior to authorizing data access, the data owner is required to present a Zero-Knowledge Proof. This proof verifies that the encrypted data is aligned with its plaintext counterpart, irrespective of the encryption scheme in use. This approach endows NuLink networks with enhanced flexibility.
 
	\section{How to Build Confidence -- Zero-Knowledge Proof in NuLink}
In this section, we will show how to use (zero-knowledge) proof/argument to ensure that all functional nodes behave in the honest way. It consists of three part: 
\begin{enumerate}
\item How to prove that the storage nodes indeed store the data?
\item How to prove that the computing nodes and proxy nodes honestly exceed the computation?
\item How to prove that the clients honestly trade with others?
\end{enumerate}

	\subsection{Proof of Storage}

 In NuLink, clients can store its data over a decentralized system. The Storage Layer of NuLink is a secure network created for the purpose of storing confidential data in encrypted form. At present, we utilize IPFS (InterPlanetary File System)\cite{ipfs} as the primary decentralized storage network. 

 The IPFS is a set of composable, peer-to-peer protocols for addressing, routing, and transferring content-addressed data in a decentralized file system. Many popular Web3 projects are built on IPFS - see the ecosystem directory (opens new window)for a list of some of these projects.

 A potential IPFS system is Filecoin\cite{filecoin}. Filecoin is a peer-to-peer network that stores files, with built-in economic incentives and cryptography to ensure files are stored reliably over time. In Filecoin, users pay to store their files on storage providers. Storage providers are computers responsible for storing files and proving they have stored them correctly over time. Anyone who wants to store their files or get paid for storing other users’ files can join Filecoin. Available storage, and the price of that storage, are not controlled by any single company. Instead, Filecoin facilitates open markets for storing and retrieving files that anyone can participate in. Filecoin is built on top of the same software powering IPFS protocol. Filecoin is different from IPFS because it has an incentive layer on top to incentivize contents to be reliably stored and accessed.

 A key components of Filecoin is its proof systems. Filecoin uses various proof of storage system\cite{ABCHKPS07} to ensure files are stored in storage nodes correctly. Here we start from the basic proof of storage system and show how it works. 

 Proof of storage system ensures data integrity and availability only at a specific time point (i.e. the time the challenge is issued). A simple example is as follows. 
 
 Suppose that a user wants to store its data $F$ in a storage node. The user can sample several random strings $\{r_i\}$, hash the concentrate of data and random strings $\{h_i=hash(F\|r_i)\}$ and store these hashes itself. Then it require the storage node to store $F$. Every time the user wants to check data integrity, it sends 
unused $r_i$ to the storage node and require the storage node to return $hash(F\|r_i)$. Easy to find, if the storage node didn't store the data $F$, it would not able to compute $hash(F\|r_i)$ for random string $r_i$. 

 However, as one can see, above solution is quite inefficient, especially at the case that the size of $F$ is huge. To solve this issue, we need to rely the idea that ``Checking that most of a file is stored is easier than checking that all is''. Divides $F$ into $k$ blocks $\{m_i\}_{i\in [k]}$. The user stores the hash tree of $\{m_i\}$.  Every time need to check data integrity, require the storage node to return blocks $\{m_{t_i}\}_{i\in [k']}$ where all $t_i$ are samples randomly from $[k]$ and $k'$ is much small than $k$. The user checks the correctness of sending blocks via hash tree.  As one can see, if the storage node drops too much blocks (e.g., half blocks), it will fails in the checking with a high probability (e.g., $1-2^{-k'}$). Here we still need to fill the gap that the storage node might drops a small number of blocks. For the case that there is only one storage node, we could use the technique of error correcting code so that dropping a small number of blocks would not fail the reconstruction of the entire files. For the case that there are lots of storage nodes, one can store the blocks simultaneously in lots of nodes to solve this issue. Further more, the proofs can be compressed via zk-SNARKs. 

 Unfortunately, tradition PoS only guarantee that a standard-alone prover or storage node had possession of some data at the time of the challenge/response. And there are various practical attacks that malicious nodes could exploit to get rewarded for storage they are not providing: Sybil attack, outsourcing attacks, generation attacks.\cite{filecoin}

\begin{itemize}
    \item Sybil Attacks: Malicious miners could pretend to store (and get paid for) more copies than the ones physically stored by creating multiple Sybil identities, but storing the data only once.
    \item Outsourcing Attacks: Malicious miners could commit to store more data than the amount they can physically store, relying on quickly fetching data from other storage providers.
    \item Generation Attacks: Malicious miners could claim to be storing a large amount of data which they are instead efficiently generating on-demand using a small program. 
\end{itemize}

In practical, we usually use two varieties of PoS, Proof-of-Replication and Proof-of-Spacetime \cite{ACET20}. 

Proof-of-Replication (PoRep) is a novel Proof-of-Storage which allows a server (i.e., the prover P) to convince a user (i.e., the verifier V) that some data D has been replicated to its own uniquely dedicated physical storage. It can prevent the Sybil attack, outsourcing attacks, and generation attacks. 

 A simple example of proof-of-replication is that, instead of storing the data blocks $\{m_i\}$, the user requires the storage nodes to store a replica $m^\tau_i$ of each data blocks. In the proof phase, the storage nodes prove the integrity and availability of replica through above PoS protocol. The key point here is that constructing a replica $m^\tau_i$ from the origin data $m_i$ and/or other replicas $m^{\tau'}_{i'}$ requires much more time than generating the PoS proofs (such a replica scheme can be constructed via seal scheme or delay encoding scheme.). And the user will reject those proofs if they takes too much time to generate. Though these methods, the malicious storage nodes cannot generate a valid proof with the help of other nodes in a short enough time. 

 Since the introduction of proof-of-replication, various of novel schemes are constructed, towards better prove/verification performance, proof size and/or security, such as \cite{Fisch19},\cite{CYHWW21} and so on.

 Proof-of-Spacetime (PoSt) schemes allow prover P to convince verifier V that P has spent some
``spacetime'' (storage space used over time) resources. A direct solution is to require the storage nodes to generate a sequence of PoRep proofs. The challenge of each proof are generated through hashing the essential challenge, loop counter and the previous proof. Using such method, the adversary cannot generate proofs parallel and it has to generate those proofs one by one and owning the replica through all the time. 

However, above solution will result a huge proof size and verification time. Since the introduce of PoSt, various of novel schemes are constructed, such as \cite{ACET20,mon23} . They reduce the proof size and verification time though various way, such as composable zk-SNARKs or verifiable delay-functions. 

In NuLink, we require a storage layer to provide storage service. We will provide two possible solutions in the future. 
 
The first solution is to rely on the existing distributed storage network, for example, Filecoin. Filecoin is built on top of IPFS protocol, and uses PoRep and PoSt to ensures data integrity and availability. We would provide a service so that NuLink users can store these data on Filecoin. After receiving a storage request from a NuLink user,  we first need to transfer user's NLK token into Filecoin cryptocurrency through existing zk-Bridge provider or other similar system we would develop and run over watch nodes. Then, we generate the storage order to store the (possibly preprocessed) data on Filecoin network. The retrieval of data is done in a similar way. 

Here, NuLink will provide the client that help the user to a) transform its NLK token into Filecoin cryptocurrency via online zk-Bridge system or watch nodes, b) (optional) encrypt the data via homomorphic encryption scheme to ensure the secrecy, c) preprocess the (encrypted) data to generate proper storage order for Filecoin, and d) generate of retrieval order for future usage of data.

 The second solution is to build the distributed storage network ourselves. On the top of IPFS protocol, we could utilize and develop the newest PoRep and PoSt schemes, reduce the communication bandwidth, and improve the prove and verification computation. We will use NLK token directly as the benefits of storage nodes.  In the NuLink network, storage nodes are required to stake NLK tokens as collateral in order to offer their services and receive corresponding rewards. If the storage nodes violated the protocol, their staked NLK tokens would be slashed as a penalty.

 We would also consider several optimizations for the goal of NuLink network, where the stored data might be as a input of some computations. For the case that the data is sensitive, the data should be encrypted via fully homomorphic encryption (FHE) scheme and using the homomorphic property of FHE to finish the computation. However, the size of FHE ciphertexts is much larger than plaintexts. One possible solution is the mixed encryption scheme, that is, encrypt the data using symmetric encryption (e.g., AES), store the symmetric ciphertexts rather than FHE, decrypt those symmetric ciphertexts homomorphically to generate the FHE ciphertexts of data for further computation via a FHE ciphertext of the secret key of symmetric encryption. However, these solution will result a lots of extra computation. 
 
 In our situation, we could use linear secret sharing scheme (with proper parameters) instead of symmetric encryption. Instead of store the data, we require the storage nodes to store a secret share of the data. In this way, unless the adversary control much enough nodes, it cannot learn any thing about the data. At the time that one need to compute over these data, we require each node to encrypt its secret share via the FHE encryption. Run the reconstruction algorithm of secret share in FHE ciphertexts and obtain the FHE encryption of the data. Due to that the reconstruction algorithm is a linear function, the computation cost is very light.

	\subsection{Zero-Knowledge proof of Computation}
 NuLink allows users outsource the computation to the computation nodes or other computation service provider. There are two main security concerns, soundness and privacy. Soundness means that the output of the computation nodes is indeed the result of the computation over these data. And privacy means that the computation nodes or computation service provider cannot learn anything about the data. For the case that the parameters of computation model is also sensitive, the privacy also requires that the parameters of computation cannot leak.  As one can see, the soundness is necessary for outsourcing computation and privacy is optional. In other words, there are several situations that privacy is not necessary. 

 No matter which situation it is, the main paradigm of the (zero-knowledge) proof of computation is as follows. Suppose that the data is $x$ (perhaps encrypted), which is sanded by the user or storage nodes, the computation function is $f$, which is public, and $r$ is the computation parameters, if exists. Generate $crs$ as the common reference string of the (zk-)SNARKs scheme for $L_f:=\{(x,y)|f(x,r)=y\}$. The computation node/service provider first computes the result of computation $y=f(x,r)$, records the computation processes and uses the prove algorithm of (zk-)SNARKs to generate the proof $\pi$ showing that $y$ is indeed the output of $f(x,r)$. After receiving the output $y$ and the proof $\pi$, the user uses the verification algorithm to verify the validness of proof $\pi$ for statement $(x,y)\in L_f$. 

 According to different cases, there will be numerous differences in practical instantiation.\smallskip

 \noindent\textbf{Case 1.} Consider the case that the privacy of data and computation parameters is not necessary. In this case, we don't need to use the zero-knowledge version of SNARKs, and we have various suitable chooses. 

 \textit{Use LPCP-based SNARKs as example.} Groth16\cite{Groth16} scheme is one of the most famous LPCP based scheme. To use it, before outsourcing computation, both parties need to transform the computation function $f$ into an arithmetic circuits $C$ such that $f(x)=y \Longleftrightarrow C(x,y)=1$. Such a process is called arithmetization. It is a deterministic process and only need to do it for once. Secondly, the user runs the setup algorithm $Setup(1^\lambda, C)$ to generate the $crs$. And it also needs to be run for only once for each computation function. The remaining process follows the above paradigm: the computation node compute the function and generate the proof, the user check the validness of proof.
 
 We need to mention here that: Usually, the setup algorithm of Groth16 scheme should be run by a trusted third party. However, in the decentralize environment, it's hard to instantiate such a party. And the correctness of crs is necessary for both soundness and zero-knowledge property of zk-SNARKs. But in this case, we do not need to consider zero-knowledge property, therefore we only need to consider the soundness. Soundness is used to protect the verifier, in this case, the user. There is no reason for users to generate a error crs to break the soundness as that it only damage themselves. Therefore, here we require the user to generate the crs.
 
 Now, we talk about the cost of using Groth16. The arithmetization of computation function and the generation of common reference string can be done in offline and only need to be done for once. Groth16 is based on a bilinear group $(\mathbb{G}_1,\mathbb{G}_2,\mathbb{G_T})$ To generate a valid proof, the prover needs to operate $O(|C|)$ group exponent operations. The proof of Groth16 only consists three group elements (two elements in $\mathbb{G}_1$ and one in $\mathbb{G}_2$). In practical, that's about 128 bytes. The verification of Groth16 is about $(|x|+|y|)$ group exponent operations and three pairing operations. In practical, that's about 2 millisecond. 

 As we can see, the performance of the proof size and verification time of Groth16 is quite good. In the meanwhile, the user needs to generate a crs for Groth16, which might cost a lot. Therefore, if the privacy is not necessary, and the user will run the function for lots of times, it is a good choice to use Groth16 as above. There are some good example about this case. For example, the meteorological observatory want to predict the tomorrow's weather from its meteorological data. In this cases, the predict function could be a public function and it will be execute for almost everyday.

\textit{Using PIOP-based SNARKs as example}. There are various PIOP-based SNARKs scheme, in NuLink, we are going to use PLONK scheme\cite{GWC19} and its developing version (e.g.,\cite{plonky2},\cite{CBBZ23}) due to their great verification and communication performance. Same as Groth16 scheme, both parties need to transform the computation function $f$ into an arithmetic circuits $C$ such that $f(x)=y \Longleftrightarrow C(x,y)=1$. Then, using existing common reference string (we will discuss it later), the prover use the computation data to generate the SNARKs proof and the user check the validness of proof. 

There are two point here we need to mentioned.

The first one is the generation of crs. For the case that the crs is a common random string, both party can use the hash of the statement or some global random string as crs. For the case that the crs is a common structure string, in PIOP-based construction, these crs are usually universal and updatabale. That is, the generation or updata of crs is independent from the proved circuit and these crs are secure as long as there are at least one honest party take part in the generation of update of these crs. Therefore we only need to generate a limited number of crs for different size circuit in NuLink and the users only need to updata these crs for only once (then the user can use these crs for even different circuit).

The second one is the preprocessing of the circuit. Although the cre can be generated independent from circuit, unlike Groth16 or other LPCP-based scheme, the verification has to be dependent with both the description of circuit and inputs. But luckily, the verification progress can be devided into two phases, offline phase and online phase. Offline phase is only (deteministically) dependent with the description of circuit and the online is only dependent with the inputs. Therefore the user can preprocess the circuit and generate the necessary information in the offline phase for each circuit and conclude the online phase efficiently. Furthermore, the necessary information are usually encoded as polynomial commitments and therefore the user can outsource this generation to computation nodes and check the correctness by open random points. 

As we discuss above, the online verification time and communication bandwidth is $O(\log s$ where $s$ is the size of circuit and by using proper scheme, the prover time is linear to circuit size. Furthermore, one can use the custom gate to further improve the performance. This approach is suit for most situations.
 
 \noindent\textit{Remark}: To use (zk-)SNARKs to generate proof, one has to known the arithmetization of the computation function (might be a arithmetic/Boolean circuit or a Rank-1 Constraint Satisfaction). The arithmetization process is a deterministic process and only need to be generate once for each function, and there are lots of stable and reliable algorithms library that help one to conclude the arithmetization quickly. Furthermore, for a widely used function and its arithmetization version, we can let third audit party to audit the correctness of arithmetization, and the user could decide whether to not to check the correctness of arithmetization itself.\smallskip

\noindent\textbf{Case 2.} Consider the case that the privacy of data is necessary but the privacy of computation parameters is not necessary. In this case, we don't need to use the zero-knowledge version of SNARKs, however, there is several differences with above case. 

In this case, in order to protect the privacy of data, NuLink will introduce fully homomorphic encryption (FHE). More detail, the user will encrypt its data and require the computation nodes to evaluate the fuction over the ciphertext via FHE. Although it seems like the this case is similar with above one except the difference of compuation fucntions, the introduce of FHE will significantly increasing the size of circuits. Now, generating the crs for Groth16 will cost a lot. Therefore it is no more suitable to choose Groth16 in this case. 

Therefore, we will mainly consider PLONK scheme in this case, especially HyperPlonk scheme which has linear prover time. Furthermore, in this case, the computation function will have lots of repeated subcircuit. That's due to that evaluate the goal function over FHE ciphertexts will contain lots of homomorphic operations and bootstrapping operations. We will use the custom gate technique of PLONK to design specific circuit for these homomorphic operations and bootstrapping operations to further reduce the size and improve the performance. \smallskip

\noindent\textbf{Case 3.} Consider the case that the privacy of computation parameters is necessary but the privacy of data is not necessary. Here we mainly consider the case that the computation is provided by a third computation provider which owns this parameters. For the case that the parameters are provided by another user and the computation node don't know such parameters, we will require this user send encrypted parameters via FHE and the node concludes the computation and proof similar to above case (we will design a better computing progress in the future). 

In case 3, the computation is as the form of $y=f(x,r)$, where $r$ is the parameter hiding from the user. Now, we need to rely on the zero-knowledge property of (zk-)SNARKs. 

Groth16 scheme: We have shown in Case 1 that the computation service provider can use Groth16 scheme to generate the proof for computation. Groth16 scheme has its zero-knowledge version, which only requires the prover to add several masking parameters. The key point here is that, in case 1 we require the user (verifier) to generate the crs for prover, as that we only consider soundness security. But here we also need to consider the zero-knowledge property, and the user might generate malicious crs to learn extra information of computation parameters and break the zero-knowledge property.

Therefore we need to check the validness of crs generated by user. There are various scheme\cite{ABL017,ALSZ21} working on it. Some of them require the user to additionally generate a proof or some auxiliary group elements for prover to check crs. This method requires the user spend significant more time for generating crs and the prover is also required to exceed a crs verification algorithm for checking crs. Other schemes develop a way that the prover can directly check the correctness of crs and don't need the verifier don't need to generate additional strings. This method doesn't increase the computation of verifier but significantly increase the work of prover. We will choose suitable method for different scenarios.

PLONK scheme and its developing varieties: Like Groth16 scheme, PLONK scheme and its developing varieties also have their zero-knowledge versions. They require the prover to add masking polynomial to hiding the witness, which will slightly increase the prover costs. And the verification algorithms remain the same as before or change slightly. Due to that these scheme are either in common randomness string model or in the universal and updatable  common reference string model, their crs can be generated honestly in NuLink and we can easily change them into their zero-knowledge version with a small performance penalty. \smallskip

\noindent\textbf{Case 4.} Consider the case that the privacy of data and computation parameters is both necessary. 

These cases can be seen as the combination of Case 2 and Case 3. In this case, in order to ensure the privacy of data, we require the user to encrypt its data via FHE and the computation service provider exceed the computation via the homomorphic property of FHE; in order to ensure the privacy of computation parameters, we will use the zero-knowledge version of the SNARKs scheme. For the same reason, we will not use Groth16 scheme for these case. As a result, we will choose the zero-knowledge version of PLONK scheme and its developing varieties in this case. The generation of proof follows the same way as Case 1. We further use the custom gate technique to reduce the size of cirucit and improve the performance, and require the prover to use masking polynomial to hiding the computation parameters. 
 
	\subsection{Zero-Knowledge proof of Transaction}

Besides ensuring the honest behaviors of storage nodes and computation nodes, zero-knowledge proof can also be used to ensure the behavior of trader is honest or the owned data satisfies the requirement. Although these scenarios can be seen as a special case of that proving the honest behaviors of computation nodes, the goal and requirement of these scenarios and the design of protocol are different. And therefore we put them into an independent subsection.

 In this subsection, we will present several scenarios of transaction and show how zero-knowledge proof works in these scenarios.\smallskip

 \noindent\textbf{Using Zero-Knowledge Proof to Choose and Buy while Hiding the Choices}
 
In NuLink, the seller might have lots of digital contents that the seller want to sell. These digital contents could be music, illustration, cartoon or games. The seller could shows the overview of its digital contents and send the full version to the one who buys it. To reduce the online communication bandwidth, the seller could 1) upload the ciphertexts of its digital contents on the storage nodes (everyone can download it) and  sends the secret keys to the one who buys it or 2) upload its digital contents in the plaintext mode on the storage nodes (only the one having admission can download it) and send a signature to the buyer which gives the buyer an admission to download the corresponding information from the storage nodes. The first method presents stronger privacy as that the storage nodes cannot learn the information of these digital contents while the second method presents skip the decryption process for buyer. Here we only consider the first mode. 

In some cases, the buyer might don't want to release its choice. It is a practicality requirement. For example, consider the following situation: A digital book seller sells books about diseases. And a patient wants to buy a book about its disease. Of course the patient doesn't want others know about what it buys as it might leakage its disease. In fact, the information of what peoples buy might leak the information about their age, sex, income, where they live and so on, which they might even not realize. 

Here we present one solution for handling above question. Suppose that the buyer wants to buy some files encrypted on the storage nodes and sold by the seller. The buyer first generates the first round of a two-round priced oblivious transfer (priced OT) with input its choice and the amount of transaction money, and broadcasts the transaction transcript (only contains the amount and the id of buyer and seller) as well as the priced OT message on the blockchain. The seller, after receives this message on blockchain, generates the second round of the priced OT with input of the secret key for decrypting files, as well as a zero-knowledge proof showing that the generation of priced OT is honest and the input secret key is consistent with the one of files. Finally, the buyer could extract these secret key from priced OT and decrypt the file. Easy to find that the details of transaction is hiding from anyone else and even the seller either doesn't know what buyer buys. And the cheater is prevent via the zero-knowledge proof.

However, in above solution, the amount of transaction money is public. It is okay the prices of the goods of seller only have a few types. If the prices of the goods are difference from each other, it is still possible to learn what the buyer buys. To avoid this issue, a direct way is to require the seller avoid has lots of prices for its goods, and a complicate solution to avoid the public of the amount.

There are two places that leak the amount, the transaction and the priced OT. For the fist one, we rely on structure of zerocoin or zerocash that, with the help of zero-knowledge proof, get rid of the leakage of amount in transaction. For the second one, we use n-out-of-2n OT instead of priced OT. For more details, let $n$ be the amount of seller's good. The buyer generates the first round message of the OT with input $r_1\| r_2$, where $r_1$ is the choice
of buying and $r_2$ is a random string which makes the Hamming weight of $r_1\| r_2$ is $n$. We additional require the buyer generate a zero-knowledge proof showing the amount of money and the amount applied by OT is consistence. And the seller exceed the same as before except that now it uses the  n-out-of-2n OT with proper inputs (the first n-th inputs are the corresponding public key and the last n-th inputs are dummy message)  instead of priced OT.


\smallskip

\noindent\textbf{Using Zero-Knowledge Proofs to Provide Value Proof while Ensuring User Privacy}

NuLink PRE api will further develop zero-knowledge proof technology to implement information filter and value proof. 
 
Specifically, when Bob wants to further filter information within a certain tag, he need to present some bonus, Alices can provide further value proof to get the rewards. The process is as follows: 

Bob sends a reward to the network, which includes tags, specific content/picture requirements, and the duration. The Filter Nodes transfer the content requirements to Alice who have files with that required tag. Within the reward duration, Alice can generate a ZKP to prove that their file contains the specific content that Bob is looking for, without exposing the rest of the file’s other contents. The Filter Node reads the file from the storage node and verifies the proof, then return a file list to bob. The bonus is distributed to users who pass the verification and to the Filter Nodes.

	\section{Future Works}

Zero-knowledge proof is a strong cryptography tools and has been wildly used in decentralized applications (like blockchains) to foster trust between clients and nodes without exposing sensitive information. In this paper, we have shown that how to use zero-knowledge proofs to 1) ensure that the files are store in the storage nodes correctly, 2) ensure that the computation of computation nodes or other nodes is correct without exposing sensitive information and 3) the transaction of several scenarios is exceed correctly without exposing extra information. 

In the future, we will first instantiate above protocols and schemes and achieve above functionalities in NuLink. And secondly, we will continue the research of the following topics:
\begin{enumerate}
\item Design new (zk-)SNARKs scheme with better prover performance to extend the supported circuit size.
\item Design new zk-rollup technique to increase throughput of NuLink.
\item Develop more functionality of transaction on NuLink.
\end{enumerate}
\newpage
\bibliography{main}
\bibliographystyle{alpha}

\end{document}